\documentclass[twocolumn,amsmath,amssymb,aps,pre,floatfix,superscriptaddress,nofootinbib]{revtex4-1}
\usepackage[utf8]{inputenc}
\usepackage{natbib}
\usepackage{graphicx}
\usepackage{float}\usepackage[tight,footnotesize]{subfigure}
\usepackage{color}

\RequirePackage{snapshot}

\begin{document}
\title{Braess Paradox in a network of totally asymmetric 
exclusion processes}
\author{Stefan Bittihn}
\affiliation{Institute for Theoretical Physics, Universit\"at zu K\"oln, 
  50937 K\"oln, Germany}
\email{bittihn@thp.uni-koeln.de, as@thp.uni-koeln.de}

\author{Andreas Schadschneider}
\affiliation{Institute for Theoretical Physics, Universit\"at zu K\"oln, 
  50937 K\"oln, Germany}

\graphicspath{{./}}

\date{\today}

\begin{abstract}
  We study the Braess paradox in the transport network as originally proposed by Braess with totally asymmetric
exclusion processes (TASEPs) on the edges. The Braess paradox describes the counterintuitive situation in which
adding an edge to a road network leads to a user optimum with higher travel times for all network users. Travel
times on the TASEPs are nonlinear in the density, and jammed states can occur due to the microscopic exclusion
principle, leading to a more realistic description of trafficlike transport on the network than in previously studied
linear macroscopic mathematical models. Furthermore, the stochastic dynamics allows us to explore the effects
of fluctuations on network performance. We observe that for low densities, the added edge leads to lower travel
times. For slightly higher densities, the Braess paradox occurs in its classical sense. At intermediate densities,
strong fluctuations in the travel times dominate the system's behavior due to links that are in a domain-wall
state. At high densities, the added link leads to lower travel times. We present a phase diagram that predicts the
system's state depending on the global density and crucial path-length ratios.
\end{abstract}

\maketitle
%%%%%%%%%%%%%%%%%%%%%%%%%%%%%%%%%%%%%%%%%%%%%%%%%%%%%%%%%%%%%%%%%%%%%%%% 
\section{Introduction}

Many problems in disciplines like physics, biology, economics and
traffic sciences can be analysed in the form of nonequilibrium
processes on networks. Examples are car traffic on road networks or
transport on biological networks like the intracellular motor protein
movement on the cytoskeleton \cite{PhysRevLett.110.098102}.
These examples, among many others, share some basic principles: the
individual building blocks or edges of the network can be described by
a transport model that retains a current in the system keeping it out
of equilibrium. A crucial step towards understanding those networks is
the investigation of relatively simple topologies. In our study of the
famous Braess paradox in a network of TASEPs we could prove the
occurrence of this effect in these networks and find some new insights
into phenomena which are of interest in the study of TASEP networks in
general.

The Braess paradox describes situations where, given that users
minimize their traveltimes selfishly, the addition of a new link
(edge) to a network does not lead to a decrease but to an increase of
traveltimes for any user or agent in the system.  As the users decide
selfishly upon their route through the network the system is said to
be in a stable state - the user optimum or Nash equilibrium - when the traveltimes are equal for all
individuals and any change of route would increase their traveltimes. It has to be distinguished from the system optimum,
which minimizes the maximum traveltime in the system and often leads to lower traveltimes~\cite{wardrop1952}. The assumption that individuals optimize their traveltimes selfishly instead of altruistically has been studied
under laboratory conditions, see e.g.~\cite{Selten2007394}.

In his original work \cite{Braess68,BraessNW05}, Braess proposed a mathematical model of a road
network without inter-edge correlations and with the traveltimes of the individual links or edges $E_i$
being linear functions $T_i(\rho_i)$ of the (average) density $\rho_i$
on the edges. He showed that for a specific choice of traveltime
functions and a specific demand (total number of agents or global
density) the paradoxical situation where the addition of an extra link can result in an user optimum
with higher traveltimes compared to the user optimum without the new
link occurs. Being originally established as an abstract mathematical
model, the effect was since shown to be a rather generic phenomenon
\cite{Steinberg83}. The regions of its occurrence were determined for specific models \cite{pasprincipio}. It was shown to exist in real-world road networks
\cite{YounGJ08} and several reports surfaced in popular literature
with actual examples, like the closure of 42nd
street in New York \cite{Kolata}.  Analogies of the effect have been
found e.g. in mechanical networks \cite{PenchinaP03} or energy
networks \cite{Witthaut2012}. For monotone traveltime functions, as demand increases, the
effect vanishes and the new route is not used anymore
\cite{Nagurney10} which is even more counterintuitive. Most previous research was done for linear cost/traveltime
functions. Recently the paradox was studied in dynamic flow models
\cite{Thunig2016946} and pedestrian dynamics~\cite{MICE:MICE12209}.

We study the effect in Braess' network with added periodic boundary conditions and totally asymmetric exclusion
processes (TASEPs) on the edges. The TASEP is the paradigmatic model for single-lane
traffic, with a traveltime function nonlinear in the density. A lot of
progress has been made in understanding how networks of TASEPs behave.
Mean-field methods have proven useful for specific networks
\cite{parmeggiani2009} and defining properties like the fundamental
diagram have been studied for simple networks \cite{parmeggiani2011}.
 
We study the system by mean field (MF) and Monte Carlo (MC) methods, classify the different states of
the network and provide examples of their characteristics. We show that our straightforward analysis breaks down in a
large intermediate density regime where fluctuations dominate the
system's behavior which can already be deduced from our MF
study of the system without the new link. Finally we present a phase
diagram of the system which shows its stationary state depending on the global density and the crucial pathlength ratios.

%%%%%%%%%%%%%%%%%%%%%%%%%%%%%%%%%%%%%%%%%%%%%%%%%%%%%%%%%%%%%%%%%%%%%%%%%%%%

\section{Model definition}

\subsection{The totally asymmetric exclusion process}

\begin{figure}[h!]
  \centering
  \includegraphics{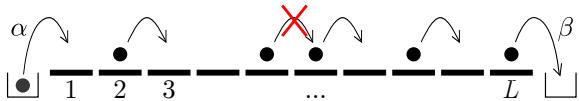}
  \caption{\label{fig:tasepeinfach}The TASEP with open boundary 
    conditions. Each of the $L$ cells can be empty or occupied by one
    particle. If a specific occupied site is chosen to be updated, the
    particle can jump forward iff the next site is empty. For the case
    of open boundary conditions, particles can enter the system at
    site 1 with probability $\alpha$ and exit the system with
    probability $\beta$ at site $L$.  For the case of periodic
    boundary conditions (not shown), site $L+1$ is identified with
    site 1 and the system becomes effectively a ring.}
\end{figure}

The TASEP is a one-dimensional cellular automaton initially introduced
as a model for protein translation \cite{BIP:BIP360060102}.  Each cell
can either be empty or occupied by one particle. The total length of a
TASEP, i.e. the number of cells, is denoted by $L$.  In the case of
periodic boundary conditions (PBC) site $L+1$ is identified with site 1. In
the case of open boundary conditions (OBC), the first site is coupled to a
reservoir which is occupied with the entrance probability $\alpha$.
The last site is connected to a reservoir which is empty with
probability $\beta$, the so-called exit probability (see
Fig.~\ref{fig:tasepeinfach}). In our study we examine the case of
random-sequential updating: a system site $i$ is chosen randomly with
the same probability for all sites. If this site is empty, nothing
happens, if the site is occupied, the particle jumps to the site $i+1$
iff site $i+1$ is empty. After $L$ of such single-site updates, one
sweep or timestep is complete.
The TASEP is well-suited for an extension to networks. Its stationary
state on a one-dimensional chain is known exactly both for periodic
boundary conditions and open boundary conditions
\cite{SchuetzD,DEHP,BlytheE07}.  In the steady state the average
density profile $\rho(i)$ does not change with time.  The
current-density relation is given by
\begin{equation}
 J(i)=\rho(i) (1-\rho(i+1)),
\end{equation}
with $J(i)$ and $\rho(i)$ being the current and the density at site
$i$, respectively. In the steady state, for a single TASEP segment the
current is independent of the site, $J(i)=J$.  For periodic boundary
conditions, the steady state of the TASEP is given by a flat density
profile, i.e. the density is site-independent with $\rho(i)=\rho=M/L$,
$M$ being the total number of particles in the system. The current is
given by $J=\rho(1-\rho)$ respectively.  For the open boundary case,
the system is also solved exactly. Here, the density is not site
independent while the current still is, as a consequence of the
continuity equation. The phase, a TASEP with open boundary conditions
will be in, depends on $\alpha$ and $\beta$.  For $L \rightarrow \infty$, the exact bulk densities
$\rho_{\text{bulk}}=\rho(L/2)$ for the different phases are given by
(see e.g. \cite{SchuetzD,DEHP})
\begin{eqnarray}
 \rho_{\text{bulk}}(\alpha,\beta)=
\begin{cases} 
\rho_{\text{LD}} = \alpha  & {\rm for\ } \alpha<\beta, \alpha<1/2 \\  
\rho_{\text{HD}} = 1-\beta & {\rm for\ } \beta<\alpha, \beta<1/2 \\ 
\rho_{\text{MC}} = 1/2     & {\rm for\ } \alpha,\beta>1/2 \\  
\rho_{\text{DW}}=1/2       & {\rm for\ } \alpha=\beta<1/2 
\end{cases}. 
\label{eq:bulkdensities}
\end{eqnarray}
with specific deviations near the boundaries in each phase~\cite{SchuetzD, DEHP}. Note that these deviations become larger, for smaller $L$. The subscript LD denotes the low density phase, HD the high density
phase, MC the maximum current phase and DW the coexistence phase or
domain wall phase\footnote{Note that for a single TASEP the latter is
  the phase transition line between HD and LD phases rather than a
  real phase itself.}. This DW phase is characterized by the diffusion
of a domain wall which separates a low density region on the left and
a high density region on the right. When measured over a long time,
the averaged density profile becomes time-independent and is given by a linear ascent from $\alpha$ to
$1-\alpha$.

For open boundary conditions the current $J$ also depends on the
entrance and exit probabilities:
\begin{eqnarray}
J(\alpha,\beta)=\begin{cases} 
\alpha(1-\alpha) & {\rm for\ } \alpha<\beta, \alpha<1/2 \\ 
\beta(1-\beta)   & {\rm for\ } \beta<\alpha, \beta<1/2 \\ 
1/4              & {\rm for\ }\alpha,\beta>1/2 
\end{cases}. 
\label{eq:currentopenbc}
\end{eqnarray}
These exact results for the one-dimensional chain can be used for
an approximate description of TASEP network dynamics.

%%%%%%%%%%%%%%%%%%%%%%%%%%%%%%%%%%%%%%%%%%%%%%%%%%%%%%%%%%%%%%%%%%%%%%%% 
\subsubsection{Traveltimes}

\begin{figure}[h!]
  \centering
  \includegraphics{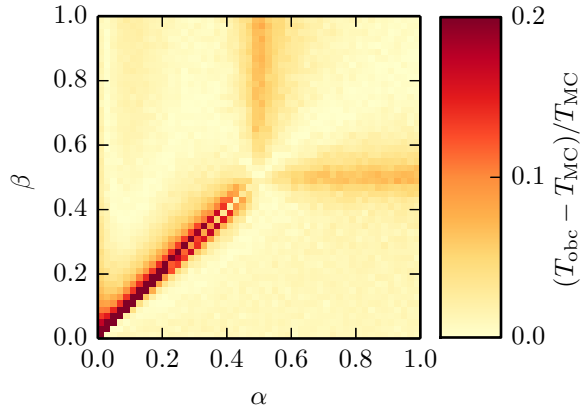}
  \caption{\label{fig:t-openbc}The relative difference of MC data 
    from the approximative traveltimes (\ref{eq:t-openbc}) for open
    boundary conditions and system size $L=100$.}
\end{figure}

Most research on the TASEP focusses on macroscopic observables like
the current $J$ or the density $\rho$ while little attention has been
given to the traveltime $T$. The traveltime is defined as the number
of timesteps a particle needs to traverse the lattice, i.e. the time
from entering site 1 until leaving site $L$.  Since the stationary
density profile is flat for periodic boundary conditions, in this case
the traveltime can be calculated exactly as
\begin{equation}
 T_{\text{PBC}}(\rho) = \frac{L}{1-\rho} .
 \label{eq:t-perbc}
\end{equation}
For this, the average velocity $v=J / \rho$ was used. Note that
the traveltime is a nonlinear function of the density in contrast to
what is usually assumed in most studies of Braess paradox.
For the case of open boundary conditions, exact traveltimes can in principle be
calculated from the exact density profiles including the exact boundary behaviour.  Eq.~(\ref{eq:t-perbc})
holds pretty well for open boundary conditions when substituting the
density by the exact bulk density $\rho=\rho_{\text{bulk}}$ of the
system given by Eq.~(\ref{eq:bulkdensities}).  From
Fig.~\ref{fig:t-openbc} we see that Eq.~(\ref{eq:t-perbc}) (with $\rho=\rho_{\text{bulk}}$) shows
notable deviations from the MC data only near the phase boundaries
and especially in the domain wall phase (i.e.\ the phase boundary
between HD and LD phases).
The latter are an effect of the fluctuations of the domain wall position in
that phase. The deviations could be minimized by measuring over a
really long time interval. However, really long measurement intervals
$[t,t+\Delta t]$ are not suitable in our case since we want to simulate
the situation of cars in a real road network where the effects for
single drivers are relevant and not the $\Delta t \rightarrow \infty$
behaviour of the system.  In this scenario, traveltimes of individual
particles, each traversing the system at a different time and thus a
different position of the domain wall and a different length of the HD
region, will depend strongly on the explicit time of measurement and
thus fluctuate from run to run.

Summarizing, we conclude that a TASEP with open boundary conditions
shows a traveltime
\begin{equation}
 T_{\text{OBC}}(\alpha,\beta) 
\approx T_{\text{PBC}}(\rho_{\text{bulk}}(\alpha,\beta)) 
\label{eq:t-openbc}
\end{equation}
with notable deviations only near the phase boundaries.
The nonlinearity of the traveltimes combined with the microscopic
exclusion principle and the stochasticity of the dynamics sets this
model apart from models previously considered in the context of the
Braess paradox. To our knowledge, so far only macroscopic models with
linear traveltimes have been studied although there are indications
for the paradox's occurrence in the nonlinear case \cite{Nagurney10}.

%%%%%%%%%%%%%%%%%%%%%%%%%%%%%%%%%%%%%%%%%%%%%%%%%%%%%%%%%%%%%%%%%%%%%%%% 
\subsection{TASEPs on networks}

Most transport processes are not limited to a single segment.
Transport takes place on networks where different routes between
origin and destination are possible. This is obviously the case for
road traffic in cities and also for freeways. The extension of the
single TASEP to networks of TASEPs has been given a lot of attention
in recent years.  Networks of TASEPs are generally not exactly
solvable.  Due to this, mean field (MF) methods and MC studies are the
tools to tackle these problems. The results obtained for the
single-link versions of the TASEP model can be used as a good starting
point to understand networks of TASEPs.

Over the years, different simple network topologies have been studied.
For the case of random sequential updates, among others, the cases of
one TASEP splitting into two lanes, then merging into one again
\cite{PhysRevE.69.066128}, two TASEPs feeding into one
\cite{1742-5468-2005-07-P07010} and all different variations of four
TASEPs \cite{parmeggiani2009} (i.e. 3 on 1, 2 on 2, 1 on 3) were
studied. Most of these cases focussed on open boundary conditions for
the whole network while also some networks with constant global
density were studied \cite{parmeggiani2009}. Beyond these simple
topologies, three general network classes, Bethe networks, Poissonian
networks and strongly correlated networks, have been examined
\cite{parmeggiani2011}. Networks with parallel update schemes
instead of the random sequential updates were examined in
\cite{PhysRevE.77.051108,Liu20094068,song2011effect}. For brief
summaries of most of the results obtained so far, see e.g.
\cite{ming2012asymmetric,1367-2630-15-8-085005}.

When studying networks it has proven most useful to explicitly
introduce so-called junction sites connecting the individual TASEPs
which form the edges of the network \cite{parmeggiani2009}.
MF treatment of networks neglects correlations between junction sites
and the neighbouring start-/endpoints of the edges. This corresponds
to treating all edges independently and neglecting inter-edge
correlations. The edges $E_i$ can then be treated as single TASEPs
with effective entrance and exit rates $\alpha^{\text{eff}}_i$,
$\beta^{\text{eff}}_i$. These effective rates are then determined by
the adjacent junction occupations. As an example consider edge $E_A$
being fed by junction $j_m$ with probability $p_{m,A}$ and exiting
into junction $j_n$. Its effective rates are then
$\alpha^{\text{eff}}_A=p_{m,A} \rho(j_m)$ and
$\beta^{\text{eff}}_A=1-\rho(j_n)$. The current in this segment is
then given by Eq.~(\ref{eq:currentopenbc}) as
$J_A=J(\alpha^{\text{eff}}_A, \beta^{\text{eff}}_A)$. To solve the MF
theory for the stationary state of a whole network, one has then to
solve the coupled particle-conservation equations for all junctions
which state that the density of a junction changes according to its
incoming minus its outgoing currents. In the stationary state this
change has to vanish. For most networks, this system of equations
can only be solved numerically. When the effective rates are known, also the traveltimes of paths through a network can be approximated by Eq.~(\ref{eq:t-openbc}). Here one has to keep in mind that the deviations from the bulk densities near the boundaries on network edges have a different form than on single TASEP segments due to the inter-link correlations. The deviations are in general larger than for single TASEPs (see e.g.~\cite{mesoon2014} for an analysis of the behaviour near the boundaries) which is why the deviations from Eq.~(\ref{eq:t-openbc}) are also larger. Still, Eq.~(\ref{eq:t-openbc}) is a good approximation of how traveltimes on network edges scale with their (bulk-) densities.

%%%%%%%%%%%%%%%%%%%%%%%%%%%%%%%%%%%%%%%%%%%%%%%%%%%%%%%%%%%%%%%%%%%%%%%% 

\subsubsection{The unbiased figure of eight network} \label{sec:figof8}

\begin{figure}[h!]
  \centering
  \includegraphics[width=0.5\columnwidth]{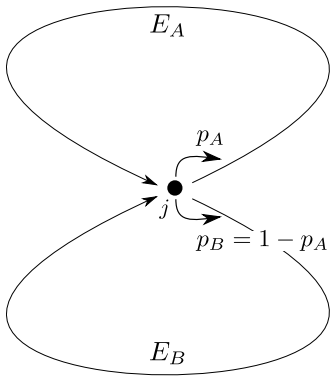}
  \caption{\label{fig:figofeight}The figure of eight 
    network consists of two symmetric TASEPs $E_A$ and $E_B$ which
    feed onto and are, with probabilities $p_A$ and $p_B=1-p_A$, fed
    by junction $j$. In the unbiased case the probabilities are equal,
    $p_A=p_B=0.5$.}
\end{figure}

\begin{figure}[h!]
  \centering
  \includegraphics{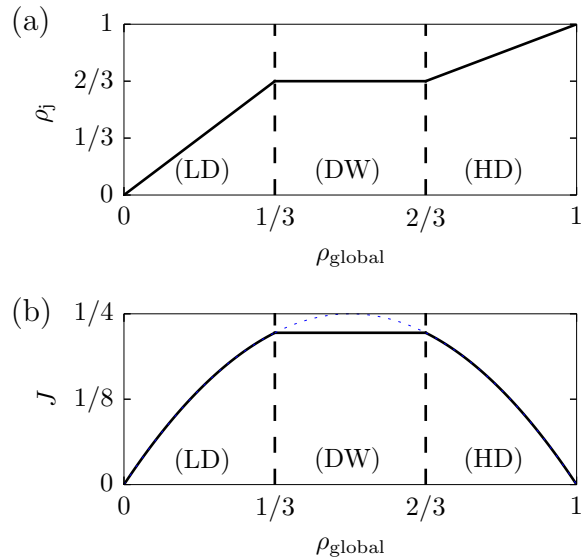}
  \caption{\label{fig:fig8graphs}The unbiased figure of eight 
    network shows a domain wall phase in a large intermediate density
    regime $1/3<\rho_{\text{global}}<2/3$. This can be seen in (a) the
    junction occupation $\rho_j$ which takes a constant value of 2/3
    in that regime according to Eq.~(\ref{eq:fig8rhoj}). As seen in (b) the current density relation shows that behaviour aswell,
    as the parabola for a single periodic TASEP $J=\rho(1-\rho)$ is
    truncated with constant value 2/9 in the DW regime according to
    Eq.~(\ref{eq:fig8j}).}
\end{figure}

Here we present the main results for a special case, the so-called
unbiased figure of eight network as shown in
Fig.~\ref{fig:figofeight}. For a more detailed treatment, see
\cite{parmeggiani2009}. The unbiased figure of eight network
consists of two TASEPs $E_A$ and $E_B$, which feed and are --
unbiasedly, i.e. with equal probability -- fed by junction $j$.
Due to the symmetry, both edges are always in the same state. This is
either a LD, HD or domain wall state. A MC phase can not be reached
since the effective entrance rates are always smaller than $1/2$ due
to the unbiased feeding. Using this symmetry, in a mean-field picture, the
particle density of the junction $\rho_j$ depends on the global
density $\rho_{\text{global}}$ as
\begin{eqnarray}
 \rho_j=\begin{cases} 2\rho_{\text{global}} & (\rho_{\text{global}}<1/3) \\ 
2/3 & (1/3<\rho_{\text{global}}<2/3) \\ 
\rho_{\text{global}} & (\rho_{\text{global}}>2/3) \end{cases} 
\label{eq:fig8rhoj}
\end{eqnarray}
and the current through the junction is given by
\begin{eqnarray}
 J=\begin{cases} 
2\rho_{\text{global}}(1-\rho_{\text{global}}) & 
(\rho_{\text{global}}<1/3) \\ 
2\cdot 2/9 & (1/3<\rho_{\text{global}}<2/3) \\ 
2\rho_{\text{global}}(1-\rho_{\text{global}}) & (\rho_{\text{global}}
>2/3) \end{cases} \label{eq:fig8j}
\end{eqnarray}
as shown in Fig.~\ref{fig:fig8graphs}.
This has an easily understandable interpretation. For low global
densities ($\rho_{\text{global}}<1/3$), both segments are in a LD
phase, while the density increases with the global density. At
$\rho_{\text{global}}=1/3$, the effective rates of the edges become
equal $\alpha^{\text{eff}}_{A/B}=\beta^{\text{eff}}_{A/B}=1/3$ which leads
to diffusing domain walls between LD and forming HD segments in
both links. The junction occupation saturates at $\rho_j=2/3$, while
the lengths of the HD regions grow with growing global density. At
$\rho_{\text{global}}=2/3$, the HD regions fill the whole edges.
This behaviour is very different to single TASEPs. In single TASEPs
with open boundary conditions domain walls only appear for fine tuned
parameters $\alpha=\beta<1/2$, while in this network, they dominate
the system over a large density regime
($1/3<\rho_{\text{global}}<2/3$) and are thus far more important
for its analysis. It has indeed been shown that all regular
networks are largely dominated by domain walls
\cite{parmeggiani2011}.

The biased version of the figure of eight network where the
probabilites for jumps from the junction to the edges are different
(Fig.~\ref{fig:figofeight}) shows two plateaus in the fundamental
diagram, corresponding to domain wall phases, for two distinct global
density regimes. For more details see \cite{parmeggiani2009}. This
study has also been extended to symmetric junctions feeding onto more
than two edges \cite{parmeggiani2011, mesoon2014}.

We have repeated the specific results for the unbiased figure of
eight here, since they correspond to a special case of our network as
discussed in Sec.~\ref{sec:MFfigof8}.

%%%%%%%%%%%%%%%%%%%%%%%%%%%%%%%%%%%%%%%%%%%%%%%%%%%%%%%%%%%%%%%%%%%%%%%% 
\subsection{Periodic Braess network}

\begin{figure}[h!]
  \centering
  \includegraphics{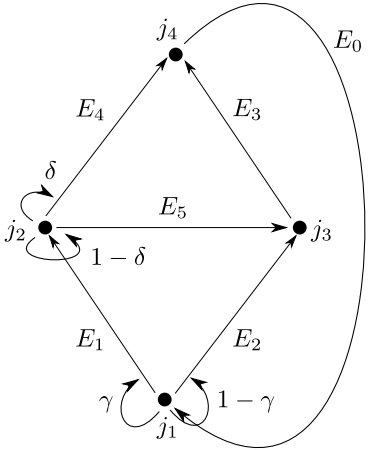}
  \caption{\label{fig:network}The network, as proposed by Braess 
    \cite{Braess68,BraessNW05}, that is studied throughout this
    paper.  Different to the original network we are considering
    periodic boundary conditions implemented by the link $E_0$. All
    links are realized by TASEPs.}
\end{figure}

We now examine the network structure, originally proposed by Braess
\cite{Braess68,BraessNW05}, shown in Fig.~\ref{fig:network}. The
individual edges $E_{1},\ldots, E_{5}$ are made up by TASEPs joined by
junction sites $j_{1},\ldots,j_{4}$. We examine the traveltimes from
start (junction $j_{1}$) to finish (junction $j_{4}$). Periodic
boundary conditions are achieved by coupling $j_{4}$ with $j_{1}$ via an
additional TASEP $E_0$ of length $L_0=1$.  Like this, the total number
of particles $M$ in the system and thus the global density
$\rho_{\text{global}}=M/(4+\sum^{5}_{i=0}L_i)$ are constant. Note that due to the random-sequential update there cannot be conflicts like two particles attempting to jump onto a junction site (like e.g. from the ends of $E_5$ and $E_2$ onto $j_3$).
Edge $E_{5}$ is considered \textit{the additional edge} which is
supposed to be added to the system. The network is always chosen to be
symmetric with
\begin{equation}
L_{1}=L_{3} \qquad\text{and}\qquad L_{2}=L_{4}\,.
%L_{1}=L_{3}=:L_{13} \text{and} L_{2}=L_{4}=:L_{24}. 
\end{equation}
We consider the case $L_{1}\leq L_{2}$. Thus, for
\begin{equation}
 L_{5}\leq L_{2}-L_{1}-1 \label{eq:l153shorter}
\end{equation}
the addition of $E_{5}$ results in a new possible route through the
system, which is of shorter or equal length as the routes without the
new link:
\begin{eqnarray}
\hat{L}_{153}&=&5 + L_{1}+ L_{3} +L_{5} \nonumber\\
&\leq& 4+L_{1}+L_{2} \label{eq:allowedl153}\\ 
&=&\hat{L}_{14}=\hat{L}_{23},
\end{eqnarray}
with $\hat{L}_{i}$ denoting lengths of routes. On junctions $j_{1}$
and $j_{2}$, the particles turn left with probabilities $\gamma$ and
$\delta$ and right with probabilites $1-\gamma$ and $1-\delta$.  For
the reminder of this paper, the system with (without) $E_5$ is also
denoted as 5link (4link) or with the superscript 5 (4), respectively.
We will use the global densities of the system with $E_5$ and without
$E_5$ when comparing the two systems. Both densities are related
through $M$ as follows:
\begin{eqnarray}
 \rho^{(5)}_{\text{global}} &=& \rho^{(4)}_{\text{global}}
\frac{ 5+2L_{1}+2L_{2} }{ 5+2L_{1}+2L_{2}+L_{5} } \nonumber\\[10pt]
 &=& \rho^{(4)}_{\text{global}}\frac{ 5+2L_{1}+2L_{2} }{ 2L_{2} 
  + \frac{\hat{L}_{153}}{\hat{L}_{14}}(4+L_{1}+L_{2}) }.
\label{eq:rho4rho5}
\end{eqnarray}
The latter equality will be used when we present the phase diagram in
Sec.~\ref{chap:phasediagram}. In our further analysis we will compare
the traveltimes of different routes through the system denoted by
$T_i$. The traveltime $T_{153}$ of route 153 is then the number of timesteps a particle sitting on
$j_1$ needs until it jumps out of $j_4$ if it traverses the system
via $E_1$, $j_2$, $E_5$, $j_3$, $E_3$. The traveltimes $T_{14}$,
$T_{23}$ of the two other possible routes are defined respectively.

%%%%%%%%%%%%%%%%%%%%%%%%%%%%%%%%%%%%%%%%%%%%%%%%%%%%%%%%%%%%%%%%%%%%%%%% 
\subsubsection{User optimum and system optimum} \label{sec:uoandso}

To determine how the new link $E_{5}$ effects the network performance
in the sense of expected traveltimes, one needs to find the new
stationary state of the system. A specific demand, given by
$\rho^{(5)}_{\text{global}}$, will result in a specific distribution
of the particles onto the three possible paths. Without traffic
regulations and with complete knowledge about expected traveltimes,
selfishly deciding drivers will choose their route through the system
such that they minimize their individual traveltimes. This results in
a stable state, the so-called user optimum (Nash equilibrium) of the
system. The user optimum ($uo$) is given by the demand distribution
that leads to equal traveltimes on all three possible routes through
the system. This state is stable since it would not make sense for
drivers to redecide for a different route if all routes have the same
traveltime and route changes would increase the traveltime. Special
cases are given if one or two routes are not used at all and have a
higher traveltime than the other used routes. One example for such a
special state would be an "all 153" state.  This state can occur for
specific demands if all particles use route 153 and nevertheless the
traveltime on this route is lower then on the unused routes 14 and 23.
This distribution would then be the user optimum if there is no other
distribution of the particles that leads to equal and lower
traveltimes on all three routes.

The other significant state is the system optimum ($so$) given by the
demand distribution which minimizes the maximum of the traveltimes on
the three routes through the system. Note that throughout the
literature there are different definitions of the system optimum.
Sometimes it is also defined as the state that minimizes the total
traveltime, i.e. the sum of all traveltimes \cite{Thunig2016946}. Here
we follow Braess \cite{Braess68,BraessNW05} who used the definition
based on the minimization of the maximum traveltime.  Furthermore the
specific choice is not as important since eventually we compare the
traveltimes of the user optima with and without the new link to deduce
how the new link effects the system. 
For our investigation the system optima are not really crucial.
However, we will later use them to distinguish different Braess
phases (see Sec.~\ref{sec-classif}).
By our definition, the maximum  traveltime $T_\text{max}(so)$ in  the
system   optimum is always  shorter  than or  equal to the traveltimes
$T(uo)$ in the   user optimum. This  means  that for the whole  system
\textit{and} all  individual  drivers,  in  terms of   traveltimes the
system optimum is always better than or equal to the user optimum.
Since the traveltimes on the three routes can be different in the
system optimum state it is not expected to be stable if drivers
decide selfishly. Drivers on a route with a higher traveltime will
redecide for a route with lower traveltime thus leading to a different
demand distribution. Without traffic regulations, the system's new
stationary stable state will be 
the user optimum.

One expects that, because of symmetry, without $E_{5}$ the
system optimum is always equal to the user
optimum at $\gamma=0.5$.  With $E_{5}$, the user optimum can be
different from the system optimum.  The Braess paradox in its original
sense is the case that $T(uo)$ with $E_{5}$ is larger than $T(uo)$
without $E_{5}$, meaning that the addition of the new route leads to a
user optimum with higher traveltimes than the user optimum without $E_5$.

%%%%%%%%%%%%%%%%%%%%%%%%%%%%%%%%%%%%%%%%%%%%%%%%%%%%%%%%%%%%%%%%%%%%%%%% 

\subsubsection{Observables} \label{sec:observables}

In our simulations the demand distributions are determined by the
turning probabilities $\gamma$ and $\delta$.  In the following a pair
$(\gamma,\delta)$, corresponding to a specific demand distribution,
will be called a strategy.  By varying $\gamma, \delta \in [0,1]$ we
realize all possible strategies.
To analyse for which values of $\gamma$ and $\delta$ the system is in
its system optimum or user optimum state, we define two observables:
\begin{eqnarray}
 \Delta T&=&|T_{14}-T_{23}|+|T_{14}-T_{153}|+|T_{23}-T_{153}| 
\label{eq:deltat} \\
 T_{\text{max}}&=&\text{max}[T_i, i\in \{14,23,153\} ]. \label{eq:tmax}
\end{eqnarray}
Analysing these observables for all possible
pairs $(\gamma,\delta)$, we can use them to infer the strategies that
correspond to the user optimum and the system optimum as:
\begin{itemize}
 \item $uo$ given by $(\gamma,\delta)_{uo}$ fulfilling $\Delta T = 0$
 \item $so$ given by $(\gamma,\delta)_{so}$ minimizing $T_{\text{max}}$. 
\end{itemize}
This is because the strategy $uo=(\gamma,\delta)_{uo}$ fulfilling $\Delta T =
0$ leads to equal traveltimes on all
routes according to (\ref{eq:deltat}).  Due to
numerical limitations and fluctuations in the simulations the case of
an exact equality is often not found, which is why we identify the
strategy corresponding to the minimium of $\Delta T$ with the user
optimum. 
Furthermore, for $\Delta T$ there are some special cases: If e.g. only
one route is used and has a lower traveltime than the unused routes,
this is defined as $\Delta T=0$, since such a state corresponds to the
user optimum as discussed in the previous subchapter. This can only
arise for the cases $\{\gamma=0 \lor \gamma=1\} \lor \{(0<\gamma<1)
\land (\delta=0 \lor \delta=1)\}$.
The strategy $so= (\gamma,\delta)_{so}$ corresponds to the state where the
longest of the three traveltimes is minimal compared to the other
strategies.

%%%%%%%%%%%%%%%%%%%%%%%%%%%%%%%%%

\subsubsection{Classification of system states}
\label{sec-classif}

Assuming that both the user optimum and the system optimum are clearly
identifiable, there is a limited number of possibilities for system
states. By "clearly identifiable" we mean that all three routes have
stable traveltimes and there is a distinct minimum for
$T_{\text{max}}$ and a distinct minimum for $\Delta T$ with a value
close to zero. In the next section we will see that this is not always
the case, since depending on $\rho_{\text{global}}$ and $\gamma$ and
$\delta$, routes or parts of routes can be in domain wall phases
leading to strongly fluctuating traveltimes. In this case there are no
stable traveltime landscapes and no clearly identifiable system or
user optima.

\begin{figure}[h!]
  \centering
  \includegraphics[width=0.99\columnwidth]{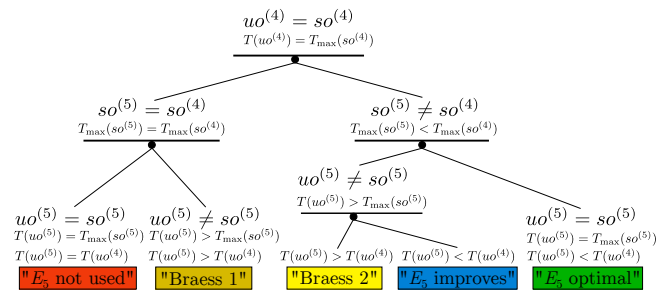}
  \caption{\label{fig:possiblestates}Tree of the possible system states, 
    if user optimum $uo$ and system optimum $so$ are clearly
    identifiable.  The superscript $(4)$ stands for the system without
    $E_{5}$, while the superscript $(5)$ denotes the system with $E_{5}$.}
\end{figure}

We now focus on the case where system optimum and user optimum are
clearly identifiable. This allows us to make a prediction about
  the phases that can be observed in the system. The possible states
are shown in Fig.~\ref{fig:possiblestates}.
If there are stable values for traveltimes and no traffic regulations,
we can build the tree of possible states from the starting point of
$uo^{(4)}=so^{(4)}$, which is expected to be always true due to symmetry.  From here, we can
compare $so^{(5)}$ and $so^{(4)}$.

If $so^{(5)}=so^{(4)}$, the additional link cannot lead to a stable state with
lower traveltimes. If in this case $uo^{(5)}=so^{(5)}$, the system will end up
in a state with $E_5$ not being used at all ("$E_5$ not used"). If $uo^{(5)}\neq so^{(5)}$, the
system will actually be in a stable state with higher traveltimes than
the system without $E_5$ since in this case $T(uo^{(5)})>T(uo^{(4)})$. This is
the classical Braess case, named "Braess 1" in this paper.

For the case $so^{(5)}\neq so^{(4)}$ the system can potentially be improved
(w.r.t. traveltimes) through the addition of $E_5$ since in this case
$T_{\text{max}}(so^{(5)})\leq T_{\text{max}}(so^{(4)})$ has to be true. If in
the 5link system $uo^{(5)}=so^{(5)}$, the system will be in its optimal state,
which has lower traveltimes than the stable state of the 4link system.
This is denoted as "$E_5$ optimal".  In the case of $uo^{(5)}\neq so^{(5)}$,
the 5link system will not be in its optimal state.  Two cases can be
distinguished. If $T(uo^{(5)})>T(uo^{(4)})$, called "Braess 2", the system
will be in a state with higher traveltimes than the 4link.  If
$T(uo^{(5)})<T(uo^{(4)})$, called "$E_5$ improves", the system will be in a state
with lower traveltimes than the 4link but still with higher
traveltimes than the system optimum of the 5link.
 
According to these possible states and additional results from the following chapters, we will present a phase diagram of
the system in Sec.~\ref{chap:phasediagram}.

%%%%%%%%%%%%%%%%%%%%%%%%%%%%%%%%%%%%%%%%%%%%%%%%%%%%%%%%%%%%%%%%%%%%%%%%%%%%
\section{Results}

First we show a mixed MF and MC study of the system without the new
road at $\gamma=0.5$. From that we can already see that, at
intermediate densities, we are not able to find stable system or user
optima since parts of the system are in domain wall phases leading to
strongly fluctuating traveltimes. In the next subsection we present
the results of MC simulations of the system with the new link.

%%%%%%%%%%%%%%%%%%%%%%%%%%%%%%%%%%%%%%%%%%%%%%%%%%%%%%%%%%%%%%%%%%%%%%%% 
\subsection{Symmetric system without the new edge} \label{sec:MFfigof8}

Here we study the system without link $E_5$. Due to symmetry, one
would expect that for all values of $\rho^ {(4)}_{\text{global}}$ the user
optimum and the system optimum were given for $\gamma=0.5$. One
expects that if on average half of the particles choose route 14 and
the other half route 23, this would lead to equal traveltimes and thus also
minimize the maximum traveltime. While this symmetry argument appears
to be obvious, it turns out that it is not true for all densities $\rho^ {(4)}_{\text{global}}$ in the sense that there are no stable
traveltimes in the system in a large intermediate global density
regime. Thus we find that we cannot use the straightforward traveltime
analysis and therefore cannot find user optima and/or system optima in
this density regime. Here we show why this is the case.

\begin{figure}[h!]
  \centering
  \includegraphics{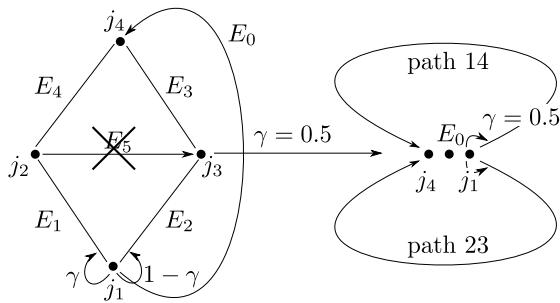}
  \caption{\label{fig:salino-periodic_figofeight} Without $E_5$ 
    and with $\gamma=0.5$ the periodic Braess network comes very close
    to the unbiased figure of eight network studied in
    chapter~\ref{sec:figof8} (for details
    see~\cite{parmeggiani2009}). The only difference is that in our
    case, the two paths are connected via three sites while
    in~\cite{parmeggiani2009} only one junction site connects the
    two paths.}
\end{figure}

\begin{figure}[h!]
  \centering
  \includegraphics{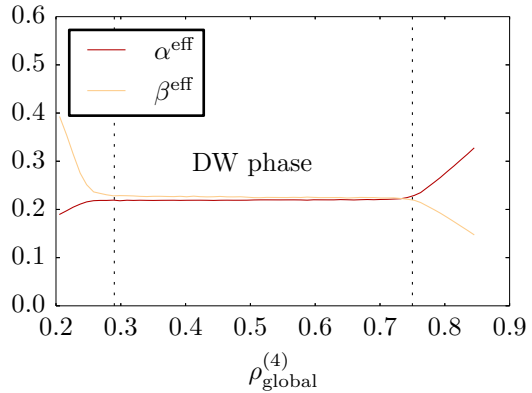}
  \vspace{-0.3cm}
  \caption{\label{fig:effrates-so4} The MC effective entrance ($\alpha^{\text{eff}}$) and 
    exit ($\beta^{\text{eff}}$) rates of paths 14 and 23 for the system without $E_5$ and
    $\gamma=0.5$. In the region of $0.29 \lesssim
    \rho^\text{(4)}_{\text{global}} \lesssim 0.75$ they are roughly equal
    and the paths will be in the domain wall phase. They were obtained
    from MC measurements of $\rho(j_1)$ and $\rho(j_4)$ for a system
    of size $L_0=1$, $L_{1}=100$, $L_{2}=500$.}
\end{figure}

\begin{figure}[h!]
  \centering
  \includegraphics{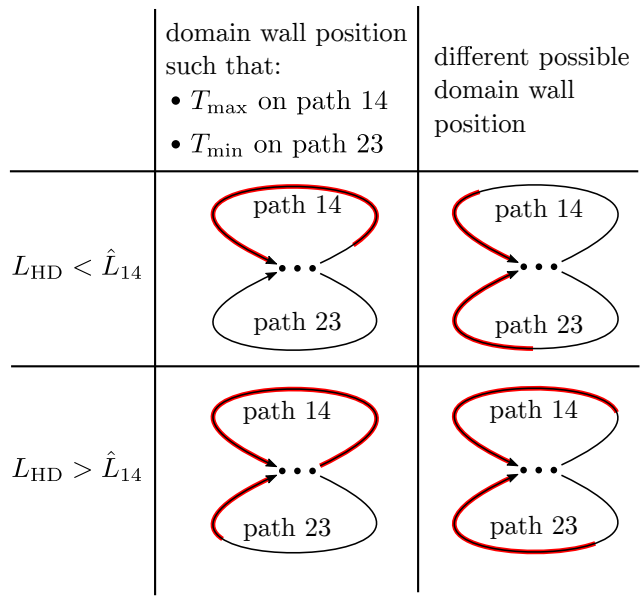}
  \caption{\label{fig:schematicminmaxtimes} Schematic of the 
      possible domain wall positions. For $L_{\text{HD}}<\hat{L}_{14}$
      (upper row), the maximum/minimum traveltimes can be measured on
      path 14/23 if the whole HD region (marked red) is on path 14
      (left column). For $L_{\text{HD}}>\hat{L}_{14}$ (lower row), the
      maximum/minimum traveltimes can be measured on path 14/23 if a
      whole path is in the HD phase, while the remnant of the HD
      region is in the other path (left column). The right column
      shows two possible different domain wall positions for the same
      $L_{\text{HD}}$ that occur at different measurement times.}
\end{figure}

\begin{figure}[h!]
  \centering
  \includegraphics{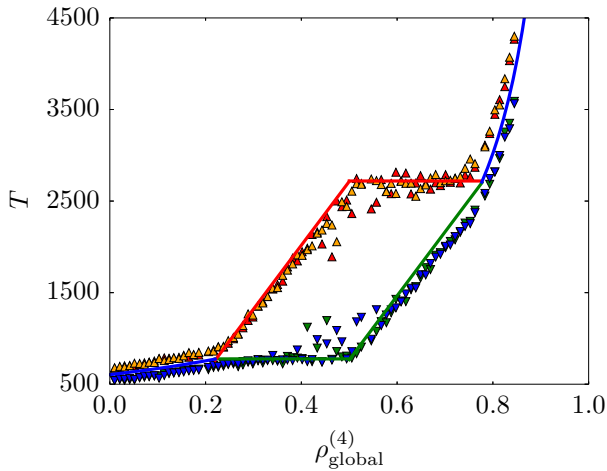}
  \caption{\label{fig:minmaxtimes} The minimum and maximum traveltimes 
      of the system without $E_5$ for $L_1=100$, $L_2=500$. In
      the large intermediate DW phase, Eq.~(\ref{eq:Tmax}) and
      (\ref{eq:Tmin}) are good approximations for the maximum (red
      line) and minimum (green line) traveltimes.  Outside of the DW
      phase, Eq.~(\ref{eq:t-openbc}) (blue lines) is a good
      approximation for the traveltimes. The equations show a good
      agreement with MC data (red/orange $\vartriangle$ for
      $T_{\text{max}, 14/23}$, green/blue $\triangledown$ for
      $T_{\text{min}, 14/23}$).  For each global density
      $\rho^{(4)}_{\text{global}}$ the traveltimes of each path were
      measured 400 times and the minimum and maximum values are
      plotted.}
\end{figure}

As seen in Fig.~\ref{fig:salino-periodic_figofeight}, without the new
link, at $\gamma=0.5$ the Braess network becomes approximately equal
to the unbiased figure of eight network. This network was studied in
\cite{parmeggiani2009} and the main results were summarized in
Sec.~\ref{sec:figof8}. The only difference is that in our case the two
edges, now given by paths 14 and 23, feed into junction $j_4$ and are
fed by junction $j_1$ while going from $j_4$ to $j_1$ via $E_0$ (of
length $L_0=1$).
Thus there are three sites connecting inputs and outputs instead of
just one junction $j$ in the original figure of eight network (see
Fig.~\ref{fig:figofeight}). While the MF arguments for the
derivation~\cite{parmeggiani2009} of the main results on the figure
of eight network, given by Eqs.~(\ref{eq:fig8rhoj}) and
(\ref{eq:fig8j}), do not hold exactly in our case, the system behaves
similarly. To visualize this, in Fig.~\ref{fig:effrates-so4} MC
measurements of the effective entrance and exit rates of paths 14 and
23 are shown.
For $\gamma=0.5$ they have the same functional dependence for both
paths and are given by
\begin{eqnarray}
 \alpha^{\text{eff}}&=&\frac{1}{2}\rho(j_1), \\
 \beta^{\text{eff}}&=&1-\rho(j_4).
\end{eqnarray}
One sees that their values are approximately equal for
$0.29\lesssim \rho^{(4)}_{\text{global}}\lesssim0.75$ and thus both paths are in
domain wall phases in this large intermediate density regime. This
region is even larger than in the former studied figure of eight
network. This is due to the fact that the three connecting sites
result in a larger effective bottleneck effect than just one junction
site. The interpretation is the same as for the figure of eight
network in Sec.~\ref{sec:figof8}. For global densities
$\rho^{(4)}_{\text{global}}\lesssim0.29$, both paths are in LD states. In the
whole intermediate density regime
$0.29\lesssim\rho^{(4)}_{\text{global}}\lesssim0.75$ both paths are in a DW state.
With growing global density, the length of the HD regions grows
compared to the LD regions. For $\rho^{(4)}_{\text{global}}\gtrsim0.75$,
both routes are in HD phases. In the DW phase we do not expect stable
traveltimes since the position of the domain wall changes constantly.
The HD regions queue behind the bottleneck (junction $j_4$), but the
position of the domain wall is changing constantly. This means that
the total length of the HD regions $L_{HD}$ is constant, but the
distribution of this region onto the two paths changes. All states
between the whole HD region being in path 14 to the whole HD region
being in path 23 are accessible. The densities of the LD and HD
regions in the domain wall phase are given by
\begin{eqnarray}
 \rho_{\text{LD}}&\approx&\alpha^{\text{eff}} \label{eq:MF1} \\
 \rho_{\text{HD}}&\approx& 1- \alpha^{\text{eff}}.\label{eq:MF2}
\end{eqnarray}
From the measurements shown in Fig.~\ref{fig:effrates-so4} we deduce
that in the whole domain wall phase, $\alpha^{\text{eff}}\approx
\beta^{\text{eff}} \approx 0.22$. From this we can now calculate the
maximum and minimum traveltimes which can be measured on both routes.
To do this, first note that the following two equations have to be
valid:
\begin{eqnarray}
  M=\rho^{(4)}_{\text{global}} L^{(4)}_{\text{tot}} &\approx &
  \rho_{\text{HD}} L_{\text{HD}}+\rho_{\text{LD}}L_{\text{LD}}\,,
  \label{eq:MF3} \\ 
L^{(4)}_{\text{tot}}& \approx &   L_{\text{HD}}+L_{\text{LD}}.
\label{eq:MF4}
\end{eqnarray}
These are not exact equalities but approximations since sharp
discontinous domain walls separating the LD and HD regions were
assumed. Furthermore we neglected the junction sites and the site of
$E_0$ to approximate the total number of sites in the system without
$E_5$ as $L^{(4)}_{\text{tot}}=2L_1+2L_2+5 \approx 2L_1+2L_2$. Using
$\rho_{\text{LD}} \approx 1 - \rho_{\text{HD}}$ from Eqs.~(\ref{eq:MF1}) and
(\ref{eq:MF2}), the system of Eqs.~(\ref{eq:MF3}) and (\ref{eq:MF4})
can be solved:
\begin{equation}
  L_{\text{HD}} = \frac{\rho^{(4)}_{\text{global}}
    L^{(4)}_{\text{tot}}}{\rho_{\text{HD}} -\rho_{\text{LD}}} -
  \frac{\rho_{\text{LD}}
    L^{(4)}_{\text{tot}}}{\rho_{\text{HD}}-\rho_{\text{LD}}}.
\end{equation}
This equation tells us how long the HD region is depending on the
global density. If we now make a further approximation and assume that
the LD and HD regions themselves have flat density profiles with a
sharp domain wall separating them, we can assume that
Eq.~(\ref{eq:t-openbc}), $T_{\text{OBC}}\approx L/(1-\rho_{\text{bulk}}
(\alpha,\beta))$, holds approximately for the
description of the traveltime on the LD and HD parts of the paths.
Using these assumptions we can then deduce the minimum and maximum
possible traveltimes of routes 14 and 23 in the DW phase:
\begin{eqnarray}
 T_{\text{max}} \approx\begin{cases} 
\frac{L_{\text{HD}}}{1-\rho_{\text{HD}}}+\frac{\hat{L}_{14}-
L_{\text{HD}}}{1-\rho_{\text{LD}}} & L_{\text{HD}}<\hat{L}_{14} \\ 
 \frac{\hat{L}_{14}}{1-\rho_{\text{HD}}} & L_{\text{HD}}>\hat{L}_{14}
\end{cases} ,
\label{eq:Tmax} \\
 T_{\text{min}}\approx\begin{cases}
\frac{\hat{L}_{14}}{1-\rho_{\text{LD}}}  & L_{\text{HD}}<\hat{L}_{14} \\
 \frac{L_{\text{HD}}-\hat{L}_{14}}{1-\rho_{\text{HD}}} 
+ \frac{L_{\text{LD}}}{1-\rho_{\text{LD}}} & L_{\text{HD}}>\hat{L}_{14}
\end{cases}  .
\label{eq:Tmin}
\end{eqnarray}
For the case where the whole HD segment is shorter than a route
($L_{\text{HD}}<\hat{L}_{14}$), the maximum traveltime is always given
if the whole HD segment is inside one route only. This leads to the
minimal traveltime on the other route since the other route is
completely in an LD phase.  The situation changes as the HD region
gets longer than a whole route, $L_{\text{HD}}>\hat{L}_{14}$.  Then
the maximum traveltime is realized if a whole route is in a HD
state which realizes the minimum traveltime the other route where the 'remnant' of the
HD segment is. These two different situations are
shown in Fig.~\ref{fig:schematicminmaxtimes}. 
This behavior and the approximative Eqs.~(\ref{eq:Tmax}) and (\ref{eq:Tmin}) were confirmed
by MC measurements as shown in Fig.~\ref{fig:minmaxtimes}.
Eqs.~(\ref{eq:Tmax}) and (\ref{eq:Tmin}) give a
very good approximation for the minimum and maximum traveltimes in the
DW phase. For the pure LD and HD phases we just assumed flat density
profiles and one stable traveltime value, approximately described by
Eq.~(\ref{eq:t-openbc}). MC measurements confirm the expected
behaviour. For each global density we made 400 individual measurements
for the traveltimes of routes 14 and 23. Then we plotted the minimum
and maximum values. The expected behaviour of a stable traveltime
value in the LD and HD regime as well as the approximate
expressions~(\ref{eq:Tmin}) and (\ref{eq:Tmax}) are confirmed.

\begin{figure}[h!]
  \centering
  \includegraphics[width=0.9\columnwidth]{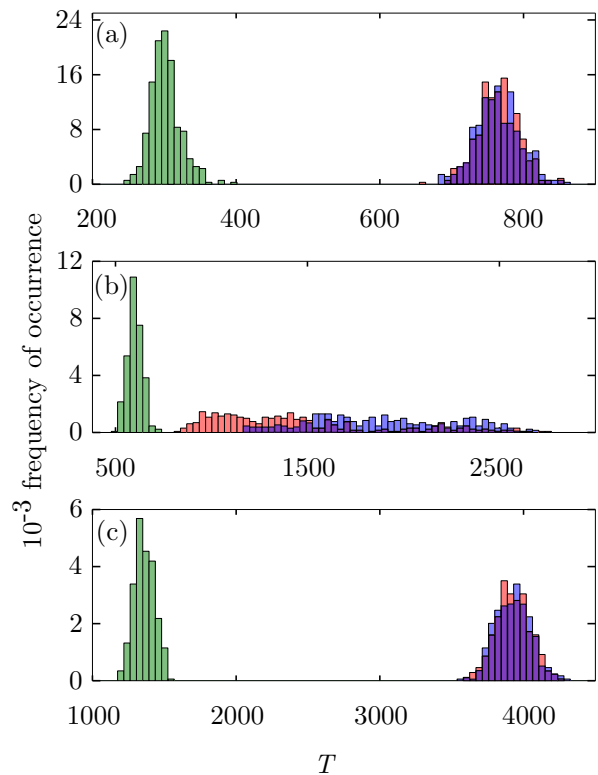}
  \caption{\label{fig:histogram-traveltimes-L5-37} Histograms 
    of the traveltime measurements in the system of size $L_0=1$, $L_{1}=100$,
    $L_{2}=500$, $L_5=37$ for $\gamma=0.5$ and $\delta=1.0$ and (a) $\rho^{(4)}_{\text{global}}=0.2$, (b) $\rho^{(4)}_{\text{global}}=0.5$, (c) $\rho^{(4)}_{\text{global}}=0.85$. The red
    bars represent the traveltimes on path 14, the blue bars those of
    path 23 and the green bars those of path 153. One can see that for
    the intermediate density, there is no well defined maximum in the
    traveltime distributions of paths 14 and 23, since these paths are
    in the domain wall phase. For each path 400 measurements were
    performed and binned.}
\end{figure}

To further clarify the effects of the fluctuating domain wall in the
DW phase we collected and binned the traveltimes of 400 individual
measurements of the traveltimes of routes 14 and 23 (traveltimes for
route 153 are included for completeness). The histograms are shown in
Fig.~\ref{fig:histogram-traveltimes-L5-37} for three different global
densities, $\rho^{(4)}_{\text{global}}\in \{0.2,0.5,0.85\}$.
We find that there is a well-defined mean in the LD and HD regions but
not in the DW region ($\rho^{(4)}_{\text{global}}=0.5$). Here all the
accessible traveltimes between $T_{\text{min}}$ and $T_{\text{max}}$
are observed with approximately the same frequency of occurrence.

The findings of these combined MF and MC arguments show that (for finite measurement intervals) in the
large intermediate density regime there are no stable expectation
values for the traveltimes of the routes in the system, even though
the system is in a nonequilibrium stationary state.
Thus it is not possible to identify the system and user optima in this
density region in the straightforward way described in
Sec.~\ref{sec:observables}.  It turns out that the system with $E_5$
is also dominated by domain walls in an even larger density regime.
Thus, with the means of traveltime measurements, we can only identify
the user and system optima of the system outside of these densities.

\subsection{Characterization of the phases}

Here we present the results of our MC simulations of the whole system
with $E_5$. The MC data were gathered as follows. The system was
always initialized randomly. Then, for given values of $(\gamma,
\delta)$, the system was relaxed/propagated for at least $2\cdot10^5$
sweeps. To measure traveltimes for the three paths through the system,
a particle was tracked on junction $j_1$ and then "manually navigated"
through the current path. During this time the rest of the system was
still propagated according to $\gamma$ and $\delta$. In detail this was done as follows: To determine e.g. a traveltime value for path 153, after relaxation, a particle sitting on $j_1$ is tagged. Since we want to measure $T_{153}$, it is then forced to jump to $E_1$, no matter the value of $\gamma$. If the tagged particle arrives on $j_2$, it is then forced to jump to $E_5$ and once it reaches $j_3$ to $E_3$. Once it reaches $j_4$, the timesteps for the whole way to get there from $j_1$ give one measurement value for $T_{153}$. During this measurement the rest of the system (all other particles) keeps evolving according to $(\gamma,\delta)$. At
least 200 individual times were measured for each path and the mean
and standard deviation were obtained. From these measurements, the
values of $\Delta T$ and $T_{\text{max}}$ were calculated. The
parameter region $\gamma, \delta \in [0.0, 1.0]$ was sweeped in steps
of 0.1. It should be noted that like this the positions of the system optima
and user optima can only be found roughly and that the exact
positions may lie between the points of the 0.1 grid. Despite the
relatively large stepwidth of 0.1 we are still able to conclude
whether system optimum and user optimum are in the same region.  This
is sufficient to deduce the phase of the system for the given
parameters. We analysed the traveltimes for different system
parameters like length ratios and densities and found that the
relevant parameters are the pathlength-ratio
$\hat{L}_{153}/\hat{L}_{14}$ (note: $\hat{L}_{14}=\hat{L}_{23}$) and
the global density $\rho_{\text{global}}$.  In this section we present
examples for how the $T_{\text{max}}$- and $\Delta T$-landscapes look
like in the different phases and also show the density profiles of the
paths in the system optima and user optima.  After that, we present a
phase diagram predicting the phase of the system dependent on the
crucial parameters.

For low global densities we find stable traveltimes and definite
minima of our observables and can thus deduce the system and user
optima of the system. For intermediate densities, fluctuations
dominate the system as already indicated in the
previous section on the system without $E_5$. For high global
densities we find stable results again but are not able to identify
system and user optima in this straightforward approach.

\subsubsection{Low global densities} 

For low global densities $\rho^{(4)}_{\text{global}}\lesssim0.29$ (and the
corresponding global densities $\rho^{(5)}_{\text{global}}$ given by
Eq.~(\ref{eq:rho4rho5})), we can find strategies
$(\gamma,\delta)$ in the parameter space where $T_{\text{max}}$ and
$\Delta T$ are minimized. This is possible since the fluctuations of
the traveltimes of the paths $i$, quantifiably by the standard
deviations $\sigma=\sqrt{\sum_i \frac{(\bar{T}-T_i)^2}{N}}$, are small and the
traveltimes have stable values.
In the low density regime, for most strategies all links of the network 
are in LD, HD or MC phases. 

\begin{figure}[h!]
  \centering
  \includegraphics[width=\columnwidth]{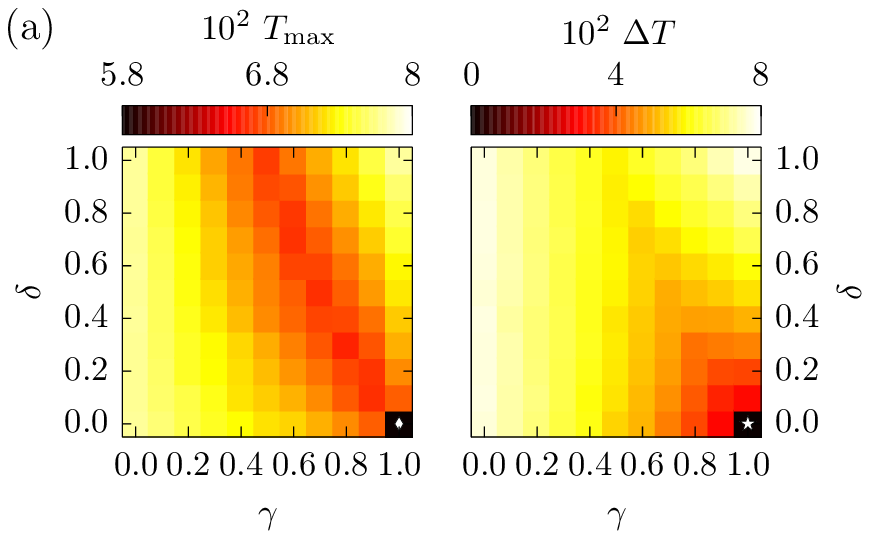}
  \includegraphics[width=0.595\columnwidth]{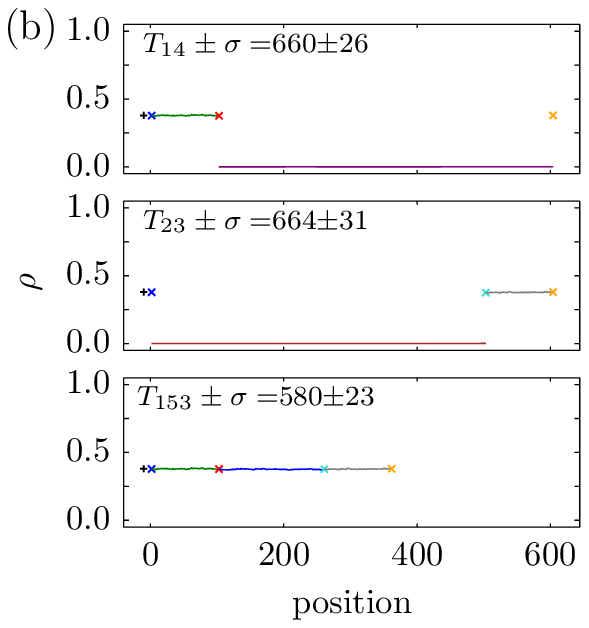}
   \includegraphics[width=\columnwidth]{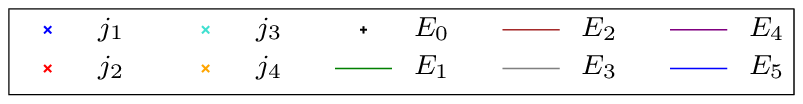}
  \caption{\label{fig:allt5} An example for the "$E_5$ optimal" 
    ("all 153") case.  Parameters are $L_0=1$, $L_{1}=100$,
    $L_{2}=500$, $L_5=157$, $M=136$. This means
    $\hat{L}_{153}/\hat{L}_{14}\approx 0.6$, $\rho^{(5)}_{\text{global}}\approx 0.1$.
    (a) $T_{\text{max}}$- and $\Delta T$-landscapes.
    The white asterisks indicate the system and user optimum, respectively.
    (b) Density profiles and average traveltimes of
    the three paths for $\text{min}(T_{\text{max}})$ and $\text{min}(\Delta T)$ - both at
    $\gamma=1.0$, $\delta=0.0$. }
\end{figure}

In Fig.~\ref{fig:allt5} an example for "$E_5$ optimal" case is shown.
This is a special "all 153" case of that phase since here both
$T_{\text{max}}$ and $\Delta T$ have their minima at
$(\gamma,\delta)_{uo}=(\gamma,\delta)_{so}=(1.0,0.0)$, meaning that
the stable state is achieved when all particles choose path 153. Thus
this is the only used path.

\begin{figure}[h]
 \centering
 \includegraphics[width=\columnwidth]{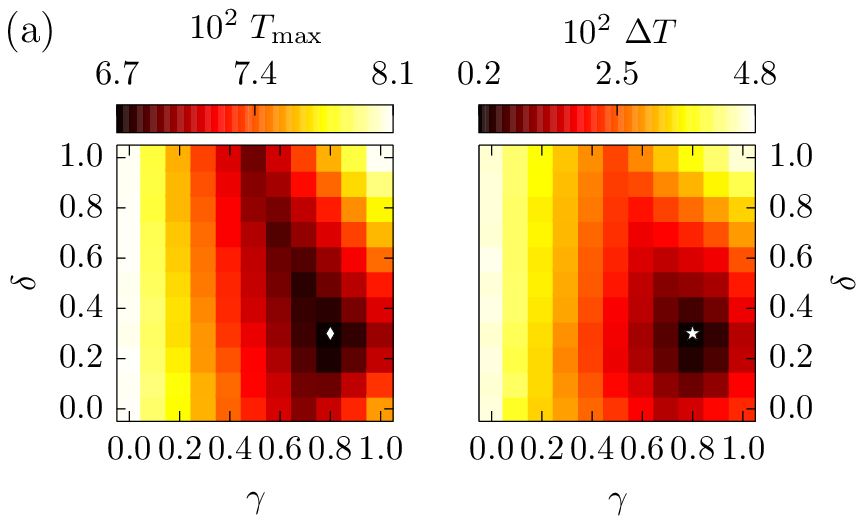}
  \includegraphics[width=0.595\columnwidth]{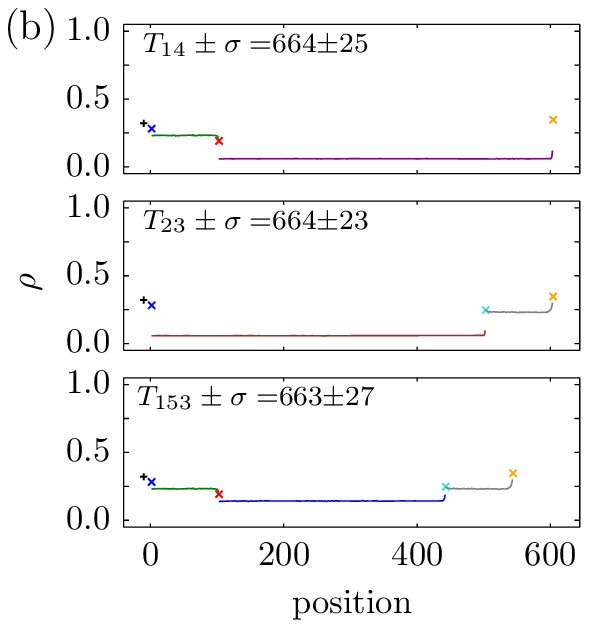}
   \includegraphics[width=\columnwidth]{./legend-densities.eps}
  \caption{\label{fig:t5useful} An example for the "$E_5$ optimal" case. 
    Parameters are $L_0=1$, $L_{1}=100$, $L_{2}=500$, $L_5=339$,
    $M=154$. This means $\hat{L}_{153}/\hat{L}_{14}\approx 0.9$,
    $\rho^{(5)}_{\text{global}}\approx 0.1$. (a) $T_{\text{max}}$- and $\Delta
    T$-landscapes.  The white asterisks indicate the system and user
      optimum, respectively. (b) Density profiles and
    average traveltimes of the three paths for
    $\text{min}(T_{\text{max}})$ and $\text{min}(\Delta T)$ - both at $\gamma=0.8$, $\delta=0.3$.}
\end{figure}

Fig.~\ref{fig:t5useful} shows an example of "$E_5$ optimal" case
where not all particles choose path 153, but the stable state
$(\gamma,\delta)_{uo}=(\gamma,\delta)_{so}$ develops at
$(\gamma,\delta)\approx(0.8,0.3)$.

\begin{figure}[h]
  \centering
  \includegraphics[width=\columnwidth]{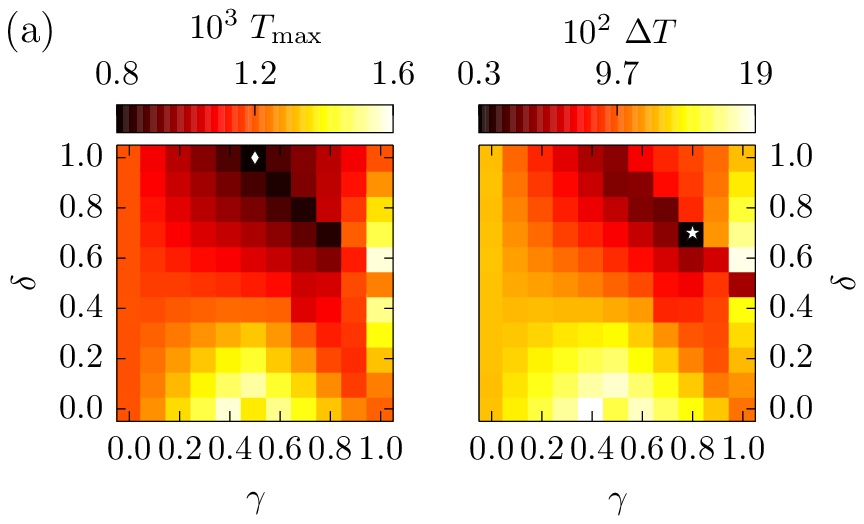}
  \includegraphics[width=0.85\columnwidth]{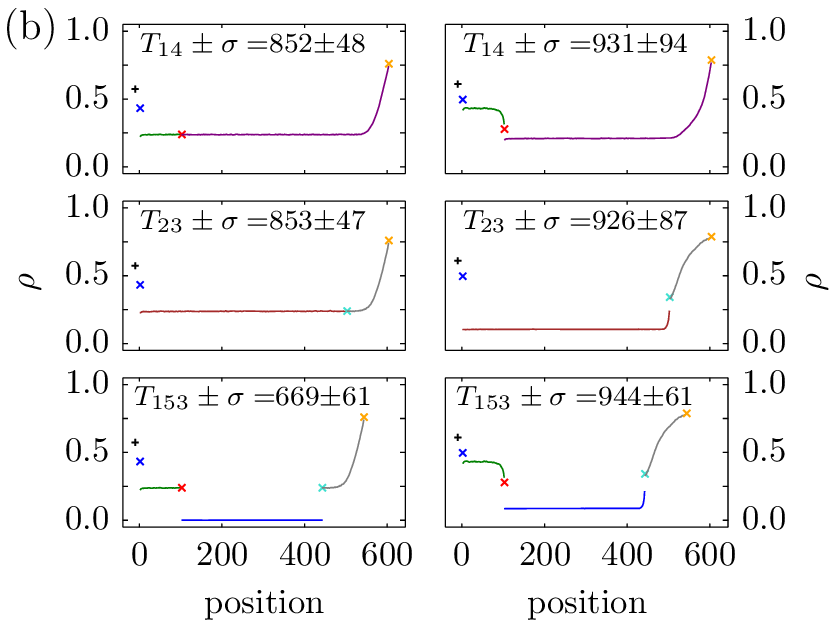}
   \includegraphics[width=\columnwidth]{./legend-densities.eps}
  \caption{\label{fig:braess1} An example for the "Braess 1" case. 
    Parameters are $L_0=1$, $L_{1}=100$, $L_{2}=500$, $L_5=339$,
    $M=309$. This means $\hat{L}_{153}/\hat{L}_{14}\approx 0.9$,
    $\rho^{(5)}_{\text{global}}\approx 0.2$.  (a) $T_{\text{max}}$- and $\Delta
    T$-landscapes.  The white asterisks indicate the system and user
      optimum, respectively. (b) Density profiles and
    average traveltimes of the three paths for $\text{min}(T_{\text{max}})$ at
    $\gamma=0.5$, $\delta=1.0$ (left) and $\text{min}(\Delta T)$ at
    $\gamma=0.8$, $\delta=0.7$ (right).}
\end{figure}

In Fig.~\ref{fig:braess1}, an example of the "Braess 1" phase is
shown.  The system optimum is at $(\gamma,\delta)_{so}=(0.5,1.0)$,
thus at the 4link system optimum, meaning that in the system optimum
$E_5$ is not used. The user optimum is found at
$(\gamma,\delta)_{uo}\approx(0.8,0.6)$. Thus, without traffic
regulations, a stable state will develop in that region resulting in
higher traveltimes for all particles, than in the 4link system. In the
density profiles shown in Fig.~\ref{fig:braess1} we can see that not
all paths are in perfect LD, HD or MC phases even in this low density
regime and that the standard deviations of the traveltimes are higher (relatively, compared to the mean value)
than in the "$E_5$ optimal" cases. This suggests that domain walls are
already present at these low global densities. We plan to investigate
this point further in the future.  Nevertheless, the fluctuations are
still small enough to consider the system to be in a stable state at
$(\gamma,\delta)_{uo}\approx(0.8,0.6)$.

%%%%%%%%%%%%%%%%%%%%%%%%%%%%%%%

\subsubsection{Intermediate global densities}

\begin{figure}[h]
  \centering
  \includegraphics[width=\columnwidth]{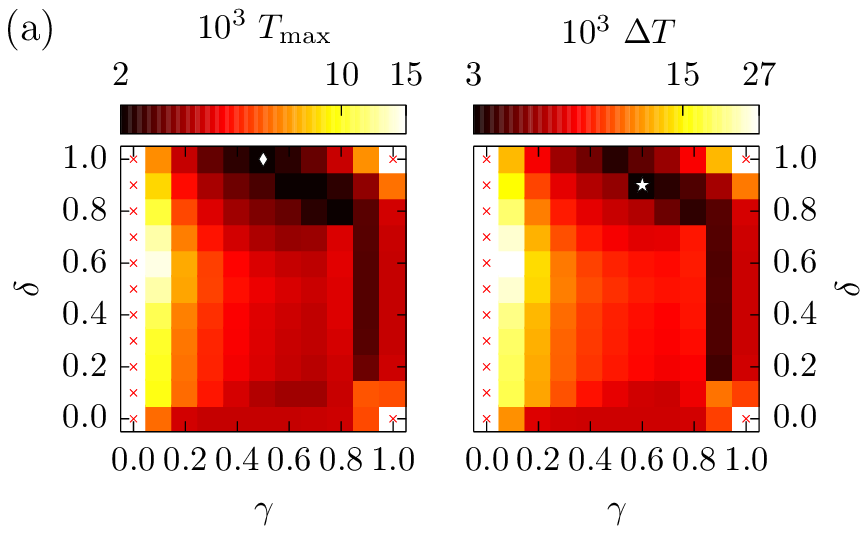}
  \includegraphics[width=0.85\columnwidth]{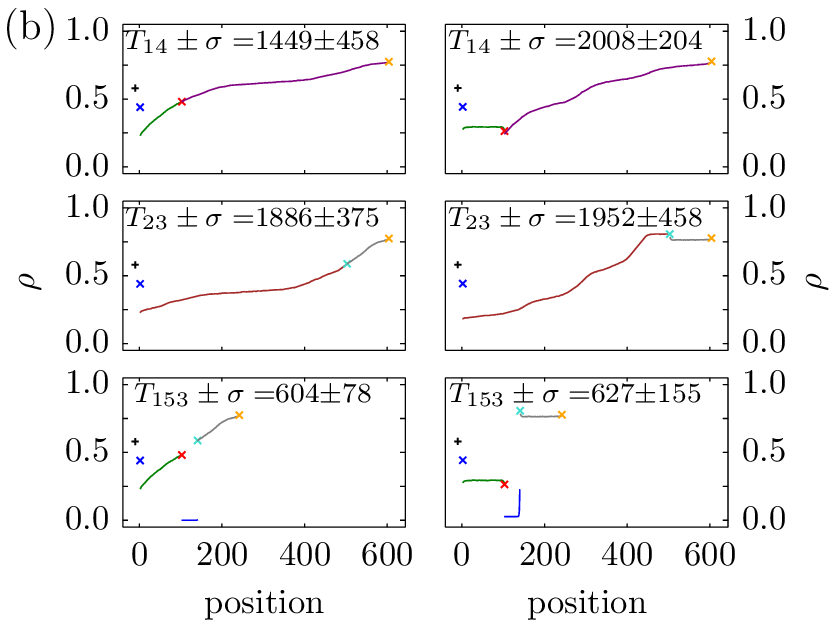}
  \includegraphics[width=\columnwidth]{./legend-densities.eps}
  \caption{\label{fig:so4flukt}  A state example from the 
    fluctuation-dominated region. Parameters are $L_0=1$, $L_{1}=100$,
    $L_{2}=500$, $L_5=37$, $M=621$. This means
    $\hat{L}_{153}/\hat{L}_{14}\approx 0.4$, $\rho^{(5)}_{\text{global}}\approx 0.5$.
    (a) $T_{\text{max}}$- and $\Delta T$-landscapes.  The
    white asterisks indicate the system and user optimum, respectively.
    Strategies marked by a red cross lead to gridlock states.
    (b) Density profiles and average traveltimes of the
    three paths for $\text{min}(T_{\text{max}})$ at $\gamma=0.5$, $\delta=1.0$
    (left) and $\text{min}(\Delta T)$ at $\gamma=0.6$, $\delta=0.9$
    (right).}
\end{figure}

For intermediate global densities
$0.29\lesssim\rho^{(4)}_{\text{global}}\lesssim0.9$, fluctuations dominate the
system with the new link for most strategies. This is an even larger
density region than for the system without the new link. These
fluctuations are due to links or whole routes of the system being in
DW states, as explained in Sec.~\ref{sec:MFfigof8}. Fluctuations are
highest for strategies close to the minima of $\Delta T$ and
$T_{\text{max}}$. An example of the $\Delta T$ -and $T_{\text{max}}$-landscapes at intermediate global densities is shown in
Fig.~\ref{fig:so4flukt}. The domain walls fingerprint can be found in
the (almost) linear average density profiles. As explained before, the
traveltimes fluctuate strongly and do not have a well-defined mean as
shown in Fig.~\ref{fig:histogram-traveltimes-L5-37}. Therefore the
$\Delta T$ -and $T_{\text{max}}$-landscapes cannot be used to determine
the system's stable state and are thus somewhat meaningless.
Nevertheless, in this whole density region the minima of
$T_{\text{max}}$ were found at $(\gamma,\delta)=(0.5,1.0)$. This
suggests that the addition of $E_5$ cannot lead to stable states with
lower traveltimes for these demands. A more detailed characterization
of this phase is another aim of future research. Here we just call
this intermediate density regime the "fluctuation-dominated phase".
Definite stable states cannot be found by the straighforward
traveltime analysis.

In certain limits, e.g.\ for $\gamma\to 0$, the traveltime diverges
due to the formation of gridlocks. In this case, on one of the routes
all sites are occupied.

%%%%%%%%%%%%%%%%%%%%%%%%%%%%%%%%%%%%%
\subsubsection{High global densities}

\begin{figure}[h]
  \centering
  \includegraphics[width=\columnwidth]{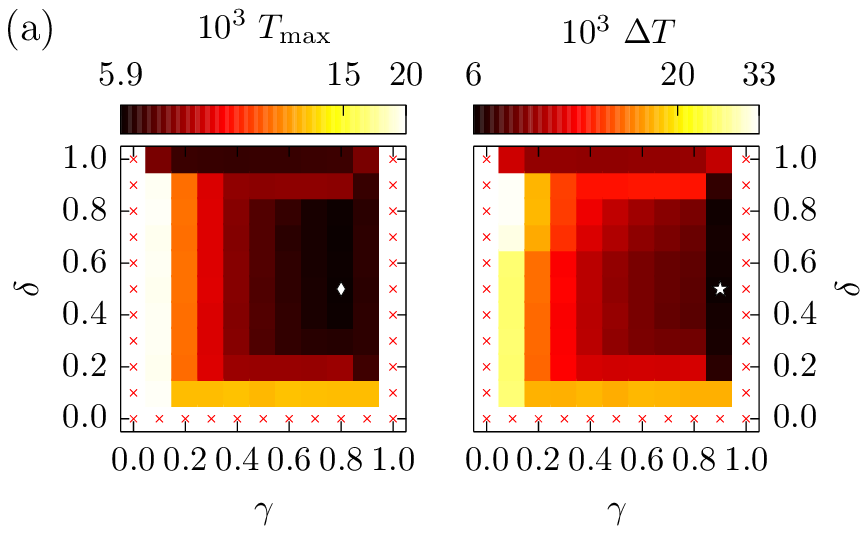}
  \includegraphics[width=0.85\columnwidth]{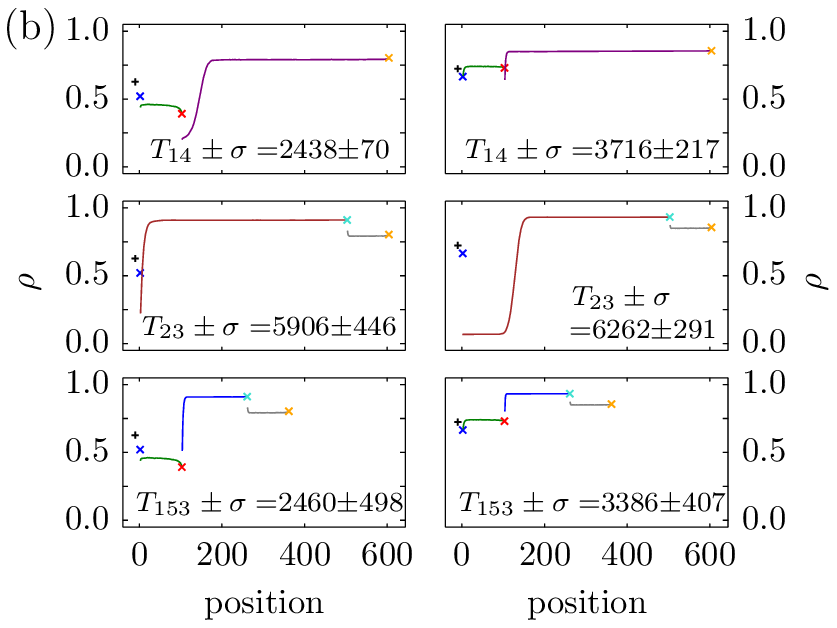}
  \includegraphics[width=\columnwidth]{./legend-densities.eps}
  \caption{\label{fig:braess3flukt} An example for the "$E_5$ improves" case. 
    Parameters are $L_0=1$, $L_{1}=100$, $L_{2}=500$, $L_5=157$,
    $M=1090$. This means $\hat{L}_{153}/\hat{L}_{14}\approx 0.6$,
    $\rho^{(5)}_{\text{global}}\approx 0.8$.  (a) $T_{\text{max}}$- and $\Delta
    T$-landscapes.  The white asterisks indicate the system and user
    optimum, respectively. Strategies marked by a red cross lead to
    gridlock states. (b) Density profiles and average
    traveltimes of the three paths for $\text{min}(T_{\text{max}})$ at
    $\gamma=0.8$, $\delta=0.5$ (left) and $\text{min}(\Delta T)$ at
    $\gamma=0.9$, $\delta=0.5$ (right).}
\end{figure}

For high densities, $\rho^{(4)}_{\text{global}}\gtrsim0.9$, $E_5$
can lead to lower traveltimes in the system. This is easily explained
by Eq.~(\ref{eq:t-perbc}), showing a diverging traveltime for
$\rho\rightarrow 1$. As seen in Fig.~\ref{fig:braess3flukt}, the 5link
system optimum moves away from the 4link system optimum.  Also, as in
the case of intermediate global densities, gridlocks can occur.  It
turns out that we cannot find a definite user optimum with
$\text{min}(\Delta T)\rightarrow0$ for these cases.  We can deduce
that $so^{(5)}\neq uo^{(5)}$ and the minimum of $\Delta T$ lies in a parameter
region where the value of $T_{\text{max}}$ is smaller than at
$(\gamma, \delta)=(0.5, 1.0)$.  This means that the system is in the
"$E_5$ improves" case according to Fig.~\ref{fig:possiblestates}.
For even higher densities
$\rho^{(4)}_{\text{global}}\geq1.0>\rho^{(5)}_{\text{global}}$,
the 4link system is full and thus we cannot compare traveltimes of the
4link with the 5link for these densities. We deduce that, trivially,
$E_5$ leads to lower traveltimes in these cases.

%%%%%%%%%%%%%%%%%%%%%%%%%%%%%%%%%%%%%%%%%%%%%%%%%%%%%%%%%%%%%%%%%%%%%%%% 

\subsection{Phase diagram} 
\label{chap:phasediagram}

\begin{figure}[h!]
  \centering
  \includegraphics{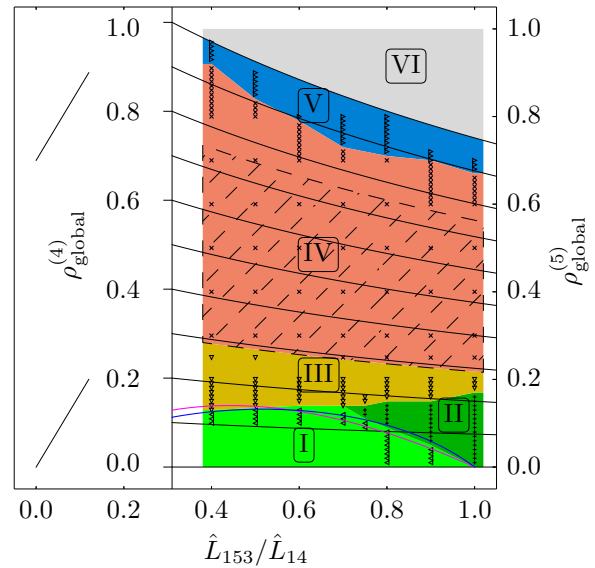}
   \caption{\label{fig:phasediagram} The phase diagram for
    $L_0=1$, $L_{1}=100$, $L_{2}=500$, as obtained by MC simulations.
    The phases of the network depend on the length ratio
    $\hat{L}_{153}/\hat{L}_{14}$ of the different paths and the global
    density $\rho^{(4/5)}_{\text{global}}$ in the system. The two
    $y$-axes show both the densities in the 4link and in the 5link
    system, related through Eq.~(\ref{eq:rho4rho5}). As we only
    consider cases where path 153 is shorter than paths 14/23, due to
    Eq.~(\ref{eq:l153shorter}) lower pathlength ratios are not
    possible here. In region~I, the system is in the "$E_5$ optimal"
    regime where all particles choose path 153 (MC-data:
    $\triangleleft$). In region~II, the system is in the general
    "$E_5$ optimal" regime (MC-data: $+$). In region~III, the system is
    in the "Braess 1" regime (MC-data: $\triangledown$). In region~IV,
    the system is in the fluctuation-dominated regime (MC-data:
    $\times$). In region~V, the system is in the "$E_5$ improves" regime
    (MC-data: $\triangleright$). In region~VI, the 4link system is
    full. The blue and magenta lines correspond to Eq.~(\ref{eq:all153-1}) and (\ref{eq:all153-2}), respectively.}
\end{figure}

In Fig.~\ref{fig:phasediagram} we present a phase diagram showing the
regimes defined in Fig.~\ref{fig:possiblestates} as a function of the
system parameters. In the presented phase diagram we used the length
ratio $L_{1}/L_{2}=1/5$, limiting the pathlength ratios to
$\hat{L}_{153}/\hat{L}_{14}\gtrsim0.34$. The measurements were
repeated for different values of that ratio and showed the same
results. The only difference is that, in accordance with
Eq.~(\ref{eq:l153shorter}), different regions of
$\hat{L}_{153}/\hat{L}_{14}$ are available. We also repeated the
measurements for a ten times larger system $\tilde{L}_i=10L_i$, to be
sure that we are not dealing with finite size effects. The larger
system also yielded the same results. Data was gathered as follows:
First the whole parameter region was sweeped in steps of 0.1. For each
pair of $\hat{L}_{153}/\hat{L}_{14}$ and
$\rho^\text{(5)}_{\text{global}}$ we analyzed the $T_{\text{max}}$ -and
$\Delta T$-landscapes like in the examples in the previous section.
After getting a rough idea of the phase landscape, the phase
boundaries were determined with finer resolution.

We can see that for low densities and small pathlength ratios the
system is in the "$E_5$ optimal"/"all 153" phase, phase I, in the sense that all
particles will use path 153 (thus $\gamma=1.0$ and $\delta=0.0$). The phase border of this phase can also be approximated analytically. Two   conditions which can be approximated by two equations have to be valid for the system to be in the "all 153" phase. Traveltime $T_{153}$ has to be lower than the other routes' traveltimes $T_{14}$ and $T_{23}$. This happens if
\begin{equation}
 T_{153} \approx \frac{\hat{L}_{153}}{1-\frac{M}{\hat{L}_{153}}} \lesssim T_{14}\approx T_{23}\approx \frac{L_{13}}{1-\frac{M}{\hat{L_{153}}}}+L_{24}, \label{eq:all153-1}
\end{equation}
since all particles choose route 153. To make sure that in this case the traveltime $T_{153}$ is actually shorter than the traveltime of the user optimum of the system without $E_5$, the second condition
\begin{equation}
 \frac{\hat{L}_{153}}{1-\frac{M}{\hat{L}_{153}}} \lesssim \frac{\hat{L}_{14}}{1-\frac{M}{2\hat{L}_{14}}}\approx \frac{\hat{L}_{23}}{1-\frac{M}{2\hat{L}_{23}}} \label{eq:all153-2}
\end{equation}
also has to hold if we assume that the stationary state of the 4link system will be reached if approximately half of the particles choose route 14 and the other half route 23. For both equations flat density profiles are assumed on the paths, which is why these equations are just approximations. The lines given by Eq.~(\ref{eq:all153-1}) (blue line) and (\ref{eq:all153-2}) (magenta line) are shown in Fig.~\ref{fig:phasediagram}. The two conditions are fulfilled in the regions of the phase diagram below both lines. This behaviour is confirmed very well by our simulations which confirm the system to be in an "all 153" state (MC-data symbols $\triangleleft$) in that region.

For larger pathlength ratios, there will also be a region where a stable state in the "$E_5$ optimal"
phase develops in which all paths are used (phase II).  For larger
densities up to $\rho^\text{(4)}_{\text{global}}\approx0.29$, the
system is in the "Braess 1" phase (phase III). Above that, for
densities $0.29\lesssim\rho^\text{(4)}_{\text{global}}\lesssim 0.9$,
the system is in the fluctuation-dominated phase (phase IV). Also the
region with strong fluctuations in the 4link system as discussed in
setion~\ref{sec:MFfigof8} is shown inside region IV (hatched area).
For $\rho^ {(4)}_{\text{global}}\gtrsim0.9$, $E_5$ leads to lower traveltimes again: Phase V
represents the "$E_5$ improves" region, while phase VI depicts the case
where the 4link system is full.

Only the boundary between phase V and phase VI is exact and the boundary of phase I is well approximated by Eq.~(\ref{eq:all153-1}) and~(\ref{eq:all153-2}). All the
others were deduced from MC data. Thus, they just represent a rough
approximation. 

Summarizing, we find that the addition of $E_5$ leads to user optima
with lower traveltimes only for really low and really high densities.
In between, its addition leads to higher traveltimes in the system.
The system is either in the classical "Braess 1" phase or fluctuations
dominate. In the intermediate density regime (phases IV and V), we
could not find the user optimum by our analysis, but we suspect that
in phase IV the addition of $E_5$ cannot result in lower traveltimes
in the system since it seems that $so^{(5)}=so^{(4)}$. In that sense, we expect that phase IV could correspond to the "$E_5$ not used" regime (see Fig.~\ref{fig:possiblestates}). The "Braess 2" phase
predicted in Sec.~\ref{sec-classif} (see Fig.~\ref{fig:possiblestates})	
could not be identified here. It remains possible that it is part of
the fluctuation-dominated regime IV.

%%%%%%%%%%%%%%%%%%%%%%%%%%%%%%%%%%%%%%%%%%%%%%%%%%%%%%%%%%%%%%%%%%%%%%%%%%%%

\section{Conclusion}

We have shown by a MC and MF analysis of traveltimes that the Braess
paradox occurs rather generically in networks of TASEPs.
Surprisingly, in a large density regime the system's behaviour is
dominated by fluctuations. This can be explained by the occurrence of
domain walls in this regime. Due to the fluctuations of the domain
wall position traveltimes can not be predicted precisely. In future
work we will study this regime further to determine the nature
of the fluctuation-dominated regime in more detail.

Away from this regime we were able to characterize the phases of the
system leading to the phase diagram shown in
Fig.~\ref{fig:phasediagram}. We have shown that the phases are
essentially determined by two relevant parameters: the global density
and the length ratio of the different paths through the network.
General arguments have predicted two different Braess phases
(Sec.~\ref{sec-classif}). In our investigation we could only verify
one of them ("Braess 1" - phase III). The occurrence of the second Braess phase ("Braess 2") could not
be established, but it could be part of the fluctuation-dominated
region IV.  Apart from the Braess phase and the fluctuation-dominated
region three phases, phase I, II and V, where the additional link indeed leads to shorter
traveltimes have been found. We could not clearly identify the user optimum in phase V.

Our results clearly show that the Braess paradox also occurs
in the presence of stochastic dynamics. Fluctuations do not surpress
the Braess phenomenon. It occurs in a relatively large subspace of
parameters and does not require fine-tuning. However, in a large 
subspace the behavior is strongly influenced by the occurrence of
fluctuating domain walls. Here the results depend on the precise
position of the domain wall. This might offer an indication of how to
control the occurrence of the paradox by controlling the dynamics
of the domain walls which could have interesting applications.

\section*{Acknowledgements}
Financial support by Deutsche Forschungsgemeinschaft (DFG) under grant SCHA~636/8-2 is gratefully acknowledged.

%%%%%%%%%%%%%%%%%%%%%%%%%%%%%%%%%%%%%%%%%%%%%%%%%%%%%%%%%%%%%%%%%%%%%%%%%%%%%% 

\bibliography{paper-braess}

%%%%%%%%%%%%%%%%%%%%%%%%%%%%%%%%%%%%%%%%%%%%%%%%%%%%%%%%%%%%%%%%%%%%%%%%%%%%%% 
\end{document}